\ifdefined\XeTeXrevision\else\ifdefined\pdfoutput\pdfoutput=1\fi
\ifdefined\pdfminorversion\pdfminorversion=7\fi\fi

\documentclass[sigconf, nonacm]{acmart}

\renewcommand\footnotetextcopyrightpermission[1]{}

\newcommand{\vldbdoi}{XX.XX/XXX.XX}
\newcommand{\vldbpages}{XXX-XXX}
\newcommand{\vldbvolume}{19}
\newcommand{\vldbissue}{12}
\newcommand{\vldbyear}{2026}
\newcommand{\vldbauthors}{Jiacheng Ding and Xiaofei Zhang}
\newcommand{\vldbtitle}{SAGA: Synthetic Agentic Graph Architecture for Temporal Benchmark Generation}
\newcommand{\vldbavailabilityurl}{https://github.com/graphuofm/SAGA}
\newcommand{\vldbpagestyle}{empty}

\usepackage{booktabs}
\usepackage{graphicx}
\usepackage{subcaption}
\usepackage{enumitem}
\usepackage{xspace}
\usepackage{amsmath}
\usepackage{algorithm}
\usepackage{algorithmic}
\usepackage{pifont}
\usepackage{times}
\usepackage{balance}   % CRC: balance columns on the last page

\graphicspath{{figures/}}

\newcommand{\sys}{\textsc{SAGA}\xspace}

\begin{document}

\title{SAGA: Synthetic Agentic Graph Architecture for\texorpdfstring{\\}{ }Temporal Benchmark Generation}

\author{Jiacheng Ding}
\email{jding2@memphis.edu}
\affiliation{%
  \institution{University of Memphis}
  \city{Memphis}
  \state{Tennessee}
  \country{USA}
}

\author{Xiaofei Zhang}
\email{xiaofei.zhang@memphis.edu}
\affiliation{%
  \institution{University of Memphis}
  \city{Memphis}
  \state{Tennessee}
  \country{USA}
}

\begin{abstract}
High quality temporal graph benchmarks with rich semantics and ground-truth anomaly labels are essential for training graph neural networks, yet remain scarce due to privacy constraints and annotation costs.
We present \sys{} (\textbf{S}ynthetic \textbf{A}gentic \textbf{G}raph \textbf{A}rchitecture), a system for generating large-scale, semantically rich temporal graphs via a four-phase pipeline.
Our \emph{Skeleton-First, Semantics-Second} architecture decouples structure from semantics: 
(\textbf{S}) an $O(1)$-per-edge skeleton generator produces power-law graphs;
(\textbf{A}) a dispatcher partitions causally ordered time blocks for parallel execution;
(\textbf{G}) LLM agents inject domain semantics using RAG-based rule bases across four domains; and
(\textbf{A}) a state alignment engine resolves conflicts via temporal replay, yielding anomaly labels as natural byproducts.
Unlike structural generators (e.g., LDBC SNB, Kronecker/R-MAT) or purely LLM-based approaches, \sys{} achieves structural realism, semantic richness, and automatic anomaly labeling in a unified framework.
On a single H100 GPU with vLLM batching, \sys{} generates 500{,}000 temporal edges with controlled anomalies in under 90 minutes, scaling to 100{,}000 nodes while maintaining clustering coefficients above 0.99.
The system supports real-time pipeline visualization, interactive multi-domain tuning (Finance/AML, Network/IDS, Cyber/APT, Transportation), and a CLI for large-scale GPU-based experiments.
\end{abstract}
%\vspace{-8ex}

\maketitle

%%% ---- PVLDB reference-format + copyright block (keep this formatting) ----
\pagestyle{\vldbpagestyle}
\begingroup\small\noindent\raggedright
\textbf{PVLDB Reference Format:}\\
\vldbauthors. \vldbtitle. PVLDB, \vldbvolume(\vldbissue): \vldbpages, \vldbyear.\\
\href{https://doi.org/\vldbdoi}{doi:\vldbdoi}
\endgroup
\begingroup\renewcommand\thefootnote{}\footnote{\noindent
This work is licensed under the Creative Commons BY-NC-ND 4.0 International License. Visit \url{https://creativecommons.org/licenses/by-nc-nd/4.0/} to view a copy of this license. For any use beyond those covered by this license, obtain permission by emailing \href{mailto:info@vldb.org}{info@vldb.org}. Copyright is held by the owner/author(s). Publication rights licensed to the VLDB Endowment.\\
Proceedings of the VLDB Endowment, Vol. \vldbvolume{} No. \vldbissue{} ISSN 2150-8097.\\
\href{https://doi.org/\vldbdoi}{doi:\vldbdoi}
}\addtocounter{footnote}{-1}\endgroup

\begingroup\small\noindent\raggedright\vspace{.2cm}
\textbf{PVLDB Artifact Availability:}\\
The source code, data, and other artifacts have been made available at \url{\vldbavailabilityurl}.
\endgroup

\section{Introduction}
\label{sec:intro}

Temporal graphs, where nodes and edges carry timestamps, underpin a wide spectrum of applications from anti-money laundering (AML)~\cite{cheng2024gnnfraud} and intrusion detection~\cite{CICIDS2017} to traffic forecasting~\cite{SUMO2018}.
Training temporal graph neural networks for these tasks requires benchmarks satisfying three properties simultaneously: \emph{structural realism} (power-law degree distributions), \emph{semantic richness} (domain-specific attributes at fine temporal granularity), and \emph{ground-truth anomaly labels} (for supervised learning without manual annotation).

Existing approaches address at most two of these aspects.
Structural generators such as Erd\H{o}s--R\'{e}nyi, Kronecker/R-MAT~\cite{erdos1960evolution,leskovec2010kronecker,chakrabarti2004rmat}, and TrillionG~\cite{park2017trilliong} scale to massive graphs but produce topology only, without semantics or labels.
Domain-specific benchmarks such as LDBC SNB and FinBench~\cite{erling2015ldbc,ldbcfinbench2022} provide realistic schemas and temporal dynamics, yet remain fixed to predefined domains and lack anomaly annotations.
Benchmark suites like TGB and TGB~2.0~\cite{huang2023tgb,gastinger2024tgb2} standardize evaluation on curated real-world datasets but do not generate new data.
Tabular generators such as CTGAN~\cite{xu2019ctgan} and LLM-based agent systems~\cite{park2023generativeagents} produce rich semantics, but fail to model graph structure, temporal dependencies, or global consistency.
%In contrast, \sys{} uniquely unifies scalable structure generation, LLM-driven domain semantics, and automatic ground-truth anomaly labeling in a single system.

\begin{table}[t]
\caption{\small Feature comparison: SAGA \textit{v.s.} others}\vspace{-2ex}
\label{tab:features}
\centering
\scriptsize
\begin{tabular}{@{}p{2.8cm}ccccc@{}}
%\begin{tabular}{@{}p{3.6cm}*{5}{>{\centering\arraybackslash}p{0.9cm}}@{}}
\toprule
\textbf{Feature} & \textbf{\sys} & \textbf{LDBC} & \textbf{AMLSim} & \textbf{Kron.} & \textbf{TGB} \\
\midrule
Power-law degree distribution          & \ding{51} & \ding{51}   & $\triangle$ & $\triangle$ & --           \\
Community structure                     & \ding{51} & \ding{51}   & \ding{55}   & $\triangle$ & --           \\
Temporal edges                          & \ding{51} & \ding{51}   & \ding{51}   & \ding{55}   & \ding{51}    \\
Configurable time granularity           & \ding{51} & $\triangle$ & \ding{55}   & \ding{55}   & \ding{55}    \\
Rich edge attributes                    & \ding{51} & $\triangle$ & \ding{51}   & \ding{55}   & --           \\
Multi-domain support                    & \ding{51} & \ding{55}   & \ding{55}   & \ding{55}   & \ding{51}    \\
Custom domain (zero-code)               & \ding{51} & \ding{55}   & \ding{55}   & \ding{55}   & \ding{55}    \\
Automatic anomaly labels                & \ding{51} & \ding{55}   & $\triangle$ & \ding{55}   & $\triangle$  \\
Real-time visualization                 & \ding{51} & \ding{55}   & \ding{55}   & \ding{55}   & \ding{55}    \\
Multi-format export                     & \ding{51} & $\triangle$ & $\triangle$ & \ding{55}   & $\triangle$  \\
\bottomrule
\end{tabular}
\vspace{0.5mm}

{\footnotesize \ding{51} = full support, $\triangle$ = partial, \ding{55} = none, -- = depends on source data.}
\end{table}

To address this gap, we present \sys (\textbf{S}ynthetic \textbf{A}gentic \textbf{G}raph \textbf{A}rchitecture), which uniquely unifies scalable structure generation, LLM-driven domain semantics, and automatic ground-truth anomaly labeling in a single system. \sys introduces a four-phase hybrid pipeline:
\textbf{S}keleton generation via igraph's C-core,
\textbf{A}gentic dispatch with causal time-block ordering,
\textbf{G}enerative injection by LLM agents with RAG rules, and
\textbf{A}lignment settlement where a global state machine transforms conflicts into labeled anomalies.
The key insight is \emph{conflict-as-feature}: agents operate with intentionally incomplete state (future balances marked \texttt{UNKNOWN}), so they naturally generate transactions violating global constraints.
The alignment engine labels these violations---overdrafts, frozen-account attempts, structuring patterns---producing ground-truth annotations without manual effort~\cite{FATF2020}. Table~\ref{tab:features} summarizes the feature comparison of SAGA against existing benchmark generation approaches.

We summarize our main contributions as follows:
(1)~A \emph{Skeleton-First, Semantics-Second} architecture combining $O(1)$-per-edge structure generation with parallel LLM semantic injection (Section~\ref{sec:system}).
(2)~A \emph{conflict-as-feature} mechanism producing ground-truth anomaly labels without manual annotation (Section~\ref{sec:system}).
(3)~Four pluggable RAG rule bases enabling domain-switching with zero code changes (Section~\ref{sec:system}).
(4)~An interactive demo with real-time pipeline visualization, multi-domain parameter tuning, and runtime event injection (Section~\ref{sec:demo}).

\section{Model and Design Foundations}
\label{sec:framework}
 
Before describing the pipeline, we formalize the three conceptual
pillars underpinning \sys: the \emph{temporal graph model},
\emph{domain rule bases}, and \emph{conflict-driven labeling}.

\noindent\textbf{Temporal Graph Model.}
\sys generates a temporal attributed graph
$G = (V, E, \mathcal{T})$, where each edge
$e_i = (u, v, t_i, \mathbf{a}_i, \ell_i)$ carries a timestamp~$t_i$,
a domain-specific attribute vector~$\mathbf{a}_i$
(amount, device, IP, risk score, \ldots), and a ground-truth label
$\ell_i \in \{\texttt{normal}\} \cup \mathcal{L}_\text{anomaly}$.
Time is organized in three user-configurable layers:
\emph{time span}~$\Delta$ (e.g., 1 year) $>$
\emph{macro-block granularity}~$B$ (e.g., 1 day, yielding
$K = \lceil\Delta/B\rceil$ blocks) $>$
\emph{micro-timestamp precision}~$\delta$ (e.g., 1 minute).
This hierarchy---directly exposed as three dropdowns in the UI
(Figure~\ref{fig:panels}d)---decouples coarse causal ordering from
fine-grained temporal realism.
The graph topology follows a power-law degree distribution via
Barab\'{a}si-Albert preferential attachment~\cite{barabasi1999emergence}
with configurable exponent $\gamma \in [1.5, 3.5]$, implemented
through igraph's C-core at $O(1)$ per edge.

\noindent\textbf{Domain Semantics \& Rule Bases.}
Domain knowledge enters \sys through pluggable RAG rule bases.
Each rule base $\mathcal{R}_\mathcal{D}$ specifies three components:
\textit{(i)}~attribute distributions
  (e.g., log-normal amounts for finance~\cite{FATF2020},
   bimodal packet sizes for network traffic~\cite{CICIDS2017});
\textit{(ii)}~temporal density functions that shape micro-timestamps
  within each block (e.g., 70\% business-hour weight for finance); and
\textit{(iii)}~validation predicates
  $\Phi = \{\phi_1, \ldots, \phi_r\}$ defining legal transactions
  (sufficient balance, active account, no self-transfer),
  where each violated predicate maps to a specific anomaly label.
Four built-in rule bases cover
\textbf{Finance/AML}~\cite{AMLSim2021,FATF2020} (12 anomaly triggers),
\textbf{Network/IDS}~\cite{CICIDS2017} (8 attack categories),
\textbf{Cyber/APT}~\cite{MITREATTACK2024} (14 ATT\&CK phases), and
\textbf{Traffic}~\cite{SUMO2018} (Krauss car-following model).
Users can also define new domains by pasting rules in natural language;
the LLM infers 10--25 typed parameters with defaults
(Figure~\ref{fig:panels}a,\,b).

\noindent\textbf{Conflict-Driven Anomaly Labeling.}
Ground-truth labels arise from \emph{validating} generated edges
against the domain rules during replay, rather than from manual annotation.
When an agent processes a task in block~$k$, its view of node~$v$ is:
\begin{equation}\label{eq:visibility}
  \hat{S}_k(v) =
  \begin{cases}
    f\!\bigl(\text{settled edges in blocks } 1..k\!-\!1\bigr)
      & \text{if settled,}\\[2pt]
    \texttt{UNKNOWN}
      & \text{otherwise.}
  \end{cases}
\end{equation}
Block-1 tasks carry known initial states; later blocks carry
\texttt{UNKNOWN} balances, forcing agents to estimate---and sometimes
violate---true constraints.
The alignment engine replays all micro-edges in global timestamp order
and evaluates each against the true state~$S_{t_i}$:
\begin{equation}\label{eq:validate}
  \ell_i =
  \begin{cases}
    \texttt{normal}            & \text{if } \forall\, \phi_j \!\in\! \Phi:\; \phi_j(e_i, S_{t_i}) = 1 \\
    \texttt{anomaly\_type}_j   & \text{if } \exists\, \phi_j:\; \phi_j(e_i, S_{t_i}) = 0
  \end{cases}
\end{equation}
Repeated violations escalate risk; exceeding $\theta_\text{freeze}$
freezes the account, causing all downstream edges to fail---producing
cascading anomaly chains that mirror real-world fraud
propagation~\cite{AMLSim2021}.
The alignment engine assigns each label \emph{type} by validating an edge
against $\Phi$ during replay; to make label \emph{volume} reproducible,
the user sets a target anomaly rate $r$, and \sys designates
$\lfloor |E|\,r \rfloor$ edges to carry violations, so the labeled-anomaly
rate is exact and deterministic regardless of LLM output variance.

%% ============================================================
\section{System Overview}
\label{sec:system}
%% ============================================================

\begin{figure}[t]
\centering
\includegraphics[width=\columnwidth]{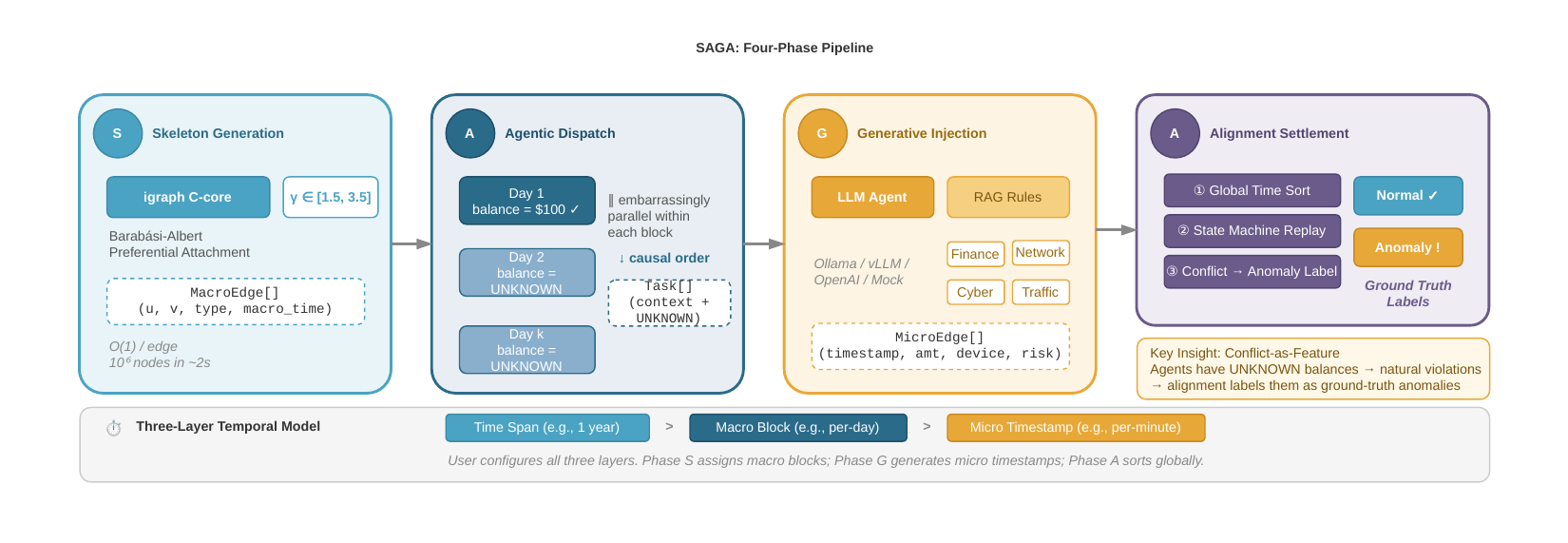}\vspace{-2ex}
\Description{Block diagram of the four-phase SAGA pipeline, from skeleton generation through agentic dispatch and generative injection to alignment settlement.}
\caption{\small The \sys pipeline. \textbf{S}: igraph C-core skeleton ($O(1)$/edge); \textbf{A$_1$}: causal time-block dispatch with \texttt{UNKNOWN} balances; \textbf{G}: LLM agents with RAG rule injection; \textbf{A$_2$}: temporal replay producing ground-truth anomaly labels.}
\label{fig:architecture}
\end{figure}

Figure~\ref{fig:architecture} illustrates the four-phase \sys pipeline.
Each phase operates on strictly typed data contracts, i.e., MacroEdge $\to$ Task $\to$ MicroEdge $\to$ FinalEdge. We now explain each phase in details. %, enabling independent optimization and testing.

\noindent\textbf{Phase S: Skeleton Generation. }
The skeleton generator produces a bare temporal graph $G_s = (V, E_s, \tau)$ where each macro-edge $e = (u, v, T, \textit{type})$ specifies endpoints and a coarse time block.
We employ igraph's C-core Barab\'{a}si-Albert preferential attachment~\cite{barabasi1999emergence} with configurable exponent $\gamma \in [1.5, 3.5]$, producing clean power-law distributions ($R^2 > 0.94$) without Kronecker artifacts~\cite{seshadhri2011indepth}.
Time is organized in three configurable layers: \emph{time span} $>$ \emph{macro block} $>$ \emph{micro timestamp}.

\noindent\textbf{Phase A$_1$: Agentic Dispatch. }
The dispatcher groups macro-edges into tasks by time block, processed in strict chronological order.
Within a block, tasks are embarrassingly parallel.
For Day\,$k$ ($k > 1$), node balances are marked \texttt{UNKNOWN}---this \emph{intentional information asymmetry} is the source of anomaly emergence.

\noindent\textbf{Phase G: Generative Injection. }
LLM agents expand macro-edges into rich micro-transactions with timestamps, amounts, device fingerprints, and risk scores.
\sys supports Ollama, vLLM~\cite{kwon2023vllm}, OpenAI API, and deterministic Mock mode via a unified \texttt{LLMCaller} interface.
Agents follow RAG-encoded temporal distributions (e.g., 70\% business-hour weight for finance).
To bound LLM hallucination, every generated edge is schema-validated and range-clamped (timestamps confined to their block, amounts to the available balance, attributes to legal enumerations); malformed outputs fall back to rule-based values, and the deterministic Mock backend reproduces structure with no LLM call.

\noindent\begingroup\emergencystretch=1.5em
\textbf{Phase A$_2$: Alignment Settlement. }
All micro-edges are sorted by timestamp and replayed through a per-node state machine tracking balance, risk level, and account status.
Each edge is validated against business rules; failed validations produce labeled anomalies 
(\texttt{anomaly\_\allowbreak overdraft}, \texttt{anomaly\_\allowbreak frozen\_\allowbreak attempt}, \texttt{anomaly\_\allowbreak structuring}).
Risk escalation creates cascading chains: repeated anomalies freeze accounts, causing all downstream transactions to be flagged---mimicking real-world fraud propagation~\cite{AMLSim2021}.\par\endgroup

\noindent\textbf{Pluggable RAG Rule Bases. }
Four rule bases provide node taxonomies, interaction patterns, temporal distributions, and anomaly conditions:
\textbf{Finance/AML}~\cite{AMLSim2021,FATF2020} (six AML patterns, twelve anomaly triggers),
\textbf{Network/IDS}~\cite{CICIDS2017} (eight attack categories),
\textbf{Cyber/APT}~\cite{MITREATTACK2024} (14 ATT\&CK tactical phases), and
\textbf{Traffic}~\cite{SUMO2018} (Krauss car-following, cascading failures).

Beyond the four built-in domains, users can define entirely new scenarios---such as IoT networks or supply chains---by pasting domain rules in natural language (Figure~\ref{fig:panels}a).
The system sends the rule text to the LLM, which automatically extracts 10--25 typed parameters (sliders, toggles, dropdowns) with sensible defaults and valid ranges (Figure~\ref{fig:panels}b).
Users retain full control: they can modify, delete, or add parameters before generation begins.
This mechanism enables domain-switching with zero code changes and zero retraining.

\section{Demonstration}
\label{sec:demo}
 
%Our demonstration runs on a React + Material Design~3 dark-theme
%dashboard with real-time WebSocket streaming and Sigma.js WebGL graph
%rendering.
%Figure~\ref{fig:overview} shows the system after a finished run, and
%Figure~\ref{fig:panels} details the interactive panels.
 
%Temporal graph benchmarks are the foundation of graph anomaly detection
%research, yet obtaining suitable data remains a significant barrier.
%Existing real-world datasets are often locked to a single domain,
%lack configurable parameters, or require extensive preprocessing before
%use.
%Researchers frequently spend more time wrangling data than advancing
%their models.
%\sys addresses this gap by enabling researchers---including those
%without domain expertise---to generate publication-ready temporal graph
%benchmarks through an intuitive visual interface, with no scripting
%required.
%We organize the demonstration around a realistic user journey across
%three progressively deeper scenarios.

\begin{figure}[t]
\centering
\includegraphics[width=\columnwidth]{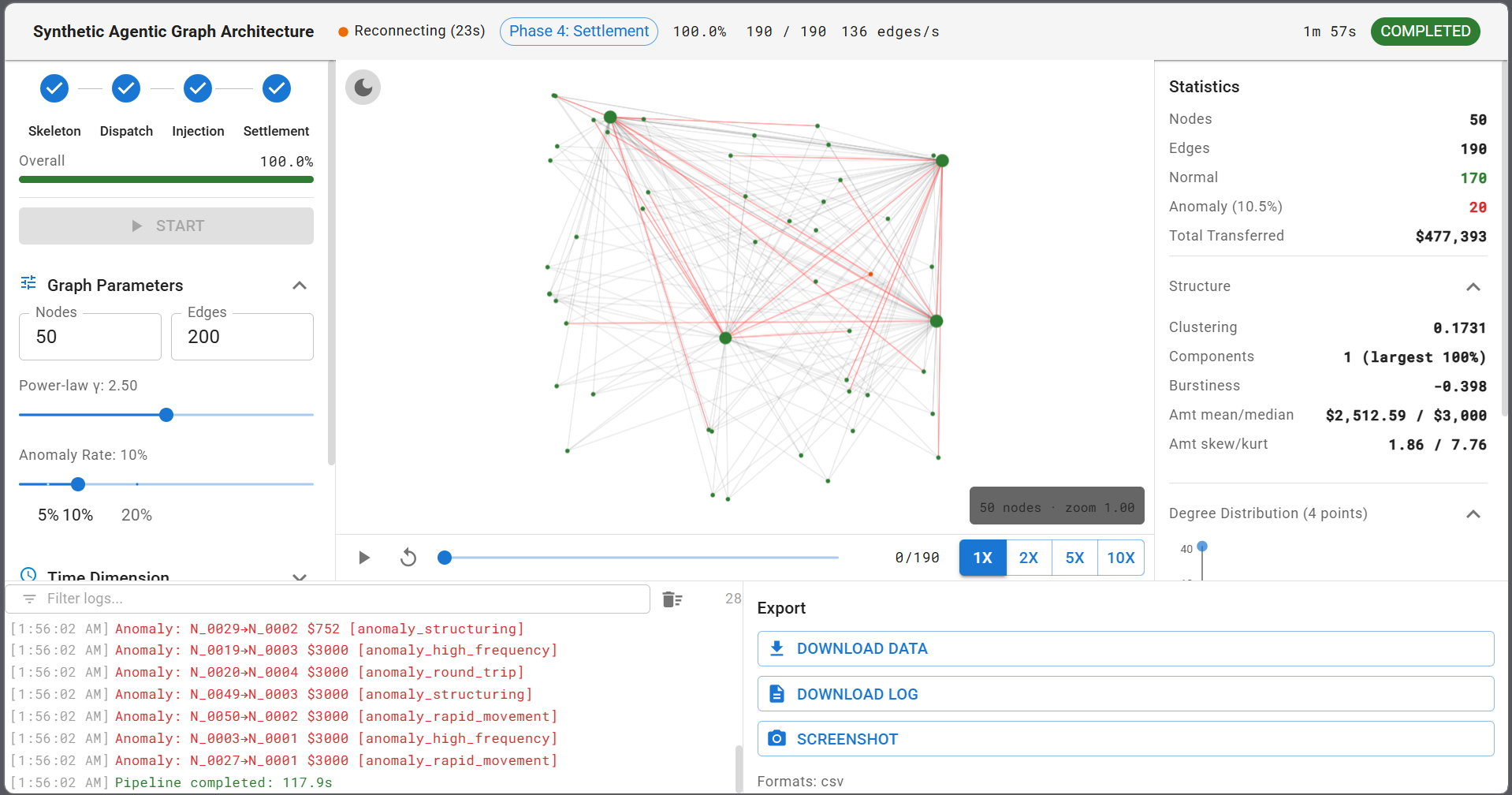}
\Description{Screenshot of the SAGA web dashboard showing the pipeline stepper, a WebGL graph view with highlighted anomalous edges, statistics panels, and a settlement event log.}
\caption{\small The \sys dashboard after a completed financial AML run
  (50 nodes, 190 edges, 10.5\% anomalies).
  Left: four-phase stepper and configuration panels.
  Center: Sigma.js WebGL graph with anomalous edges highlighted.
  Right: real-time statistics.
  Bottom: settlement event log with color-coded anomaly detections.}
\label{fig:overview}
\end{figure}

\begin{figure}[t]
\centering
\includegraphics[width=\columnwidth]{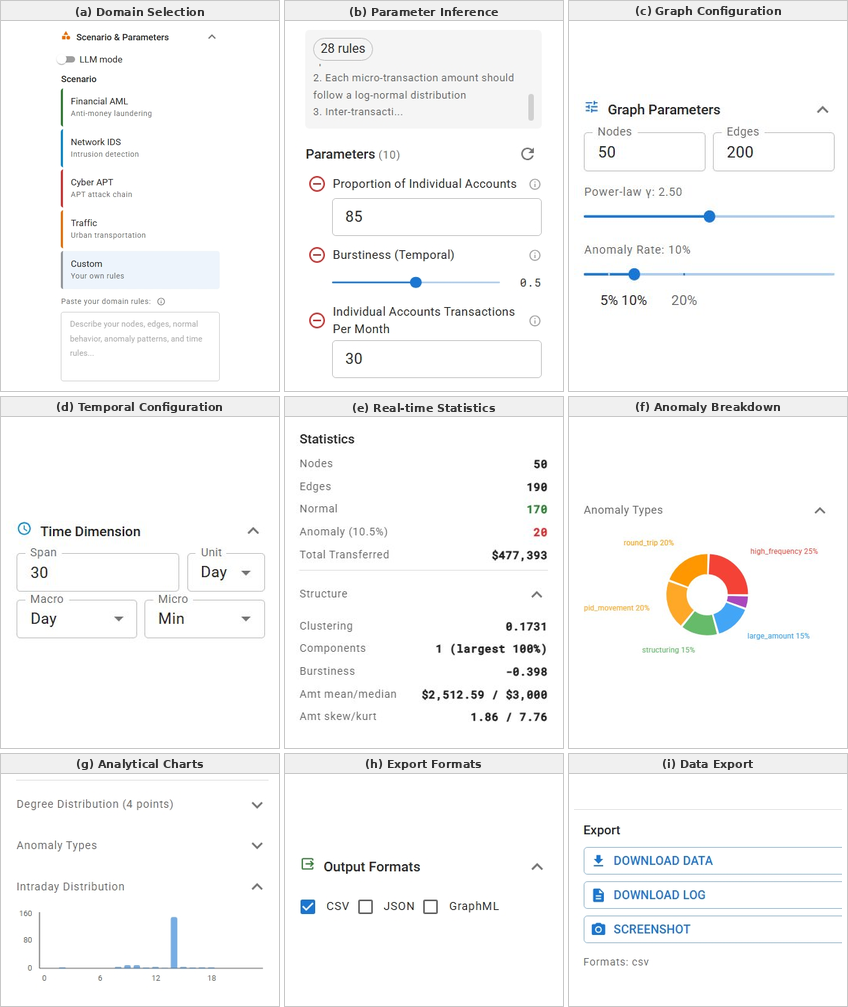}
\Description{A three-by-three grid of screenshots of SAGA's interactive control panels for domain selection, parameter tuning, topology and temporal configuration, statistics, anomaly breakdown, and export options.}
\caption{\small Interactive control panels:
  domain selection and custom rules~(a),
  LLM-inferred typed parameters~(b),
  graph topology configuration~(c),
  three-layer temporal model~(d),
  real-time statistics~(e),
  anomaly type breakdown~(f),
  intraday temporal distribution~(g),
  export format selection~(h,\,i).}
\label{fig:panels}
\end{figure}

Our demonstration is implemented as a React + Material Design~3 dashboard with real-time WebSocket streaming and Sigma.js WebGL rendering.
Figure~\ref{fig:overview} shows the system after a completed run, while Figure~\ref{fig:panels} presents the interactive control panels.

We organize the demo around a realistic user workflow, guiding participants through three progressively deeper scenarios that illustrate \sys's capabilities from rapid prototyping to large-scale deployment.

\noindent\textbf{Scenario 1: Quick Start --- From Zero to a Benchmark
in Minutes.}
A researcher needs an AML transaction dataset but finds
that existing benchmarks (e.g., Elliptic Bitcoin~\cite{ellipticbitcoin2019})
are fixed in scale and lack configurable anomaly rates.
She opens the \sys web dashboard, selects \emph{Financial AML}
from the scenario panel (Figure~\ref{fig:panels}a), and the system
instantly loads 28 domain rules and infers 10 typed parameters---sliders,
toggles, and dropdowns---with sensible defaults
(Figure~\ref{fig:panels}b).
She adjusts the graph to 50 nodes and 200 edges for a quick pilot run
(Figure~\ref{fig:panels}c), sets the time span to 7 days at
day/minute granularity (Figure~\ref{fig:panels}d), and clicks
\emph{Start}.
The four-phase stepper animates in real time: skeleton edges appear in
the graph canvas within a second, tasks dispatch by time block, LLM
agents stream micro-edges with live speed counters, and settlement
results flash green (normal) or red (anomaly) in the event log.
The entire run completes in under 10 seconds in Mock mode.
She inspects the statistics panel (Figure~\ref{fig:panels}e): 190
edges, 20 anomalies (10.5\%), clustering coefficient 0.17.
She drags the anomaly rate slider from 10\% to 25\% and reruns---the
anomaly breakdown pie chart (Figure~\ref{fig:panels}f) updates
immediately, confirming precise control.
Satisfied with the pilot, she clicks \emph{Download Data}
(Figure~\ref{fig:panels}h,\,i) and receives clean CSV files, where edges
with timestamps, amounts, labels, and a \texttt{saga\_meta.json}
recording all parameters are ready to be loaded into PyG or DGL without
any preprocessing.

\noindent\textbf{Scenario 2: Domain Switching --- Exploring Multiple
Industries Without Code.}
Assume the same researcher now needs a network intrusion dataset for a
comparative study.
Rather than searching for a new benchmark, she switches the domain
dropdown to \emph{Network IDS} (Figure~\ref{fig:panels}a).
The parameter panel regenerates automatically: \emph{Proportion of
Individual Accounts} becomes \emph{Ratio of Internal Hosts},
\emph{Burstiness} shifts to reflect diurnal packet patterns, and
new parameters such as \emph{Average Packet Size} appear---all
inferred by the LLM from the IDS rule base~\cite{CICIDS2017}.
No code changes, no configuration files, no domain expertise required.
She can even paste custom rules in natural language (e.g.,
describing an IoT sensor network) and the system generates a
fully typed parameter panel in seconds.
This zero-code domain switching is what distinguishes \sys from
fixed-schema generators like LDBC SNB, which require schema redesign
and code modifications for each new domain.

\noindent\textbf{Scenario 3: Scaling Up --- From Prototype to
Large-Scale Experiments.}
Having validated her approach on small graphs, the researcher
scales to 10{,}000 nodes and 50{,}000 edges.
The web dashboard remains interactive: Sigma.js renders the graph at
60fps via WebGL, and the time-axis scrubber enables replay at
1$\times$/2$\times$/5$\times$/10$\times$ speed.
For even larger experiments (100K+ nodes), she switches to the
\sys CLI, which accepts the same parameters via YAML configuration
files and runs on GPU clusters.
On a single H100 with vLLM continuous batching (256 concurrent
requests), \sys generates 50{,}000 edges in 9 minutes and 500{,}000
edges in 89 minutes, scaling linearly with edge count.
A deterministic Mock mode bypasses LLM calls entirely, producing
500K edges with full structural annotations in 30 seconds for rapid
prototyping and CI testing.
Throughout, the exported data maintains consistent quality:
power-law $R^2 > 0.94$, clustering coefficients above 0.99 at
scale, and exact anomaly-rate control from 0\% to 30\%.
Each export includes multi-format outputs (CSV, JSON, GraphML)
directly loadable by Pandas, Neo4j, and igraph, plus a metadata
file for full reproducibility.
\sys shifts the researcher's effort from \emph{data acquisition} to \emph{data specification}---freeing them to focus on mining rather than curation.

\noindent\textbf{Implementation.}
The backend is built on Python~3.10+ with python-igraph, asyncio,
orjson, and numpy.
The frontend uses React~18, Vite, MUI~v5, Sigma.js, and Recharts.
\sys supports four LLM backends (Ollama, vLLM~\cite{kwon2023vllm},
OpenAI API, Mock) via a unified \texttt{LLMCaller} interface,
switchable through a single environment variable.
The system has been tested on NVIDIA Quadro RTX~6000 (24\,GB) and
H100 (80\,GB).

\section{Evaluation}
\label{sec:eval}
 
We evaluate \sys on \textit{structural fidelity}, \textit{controllability}, and \textit{scalability} using an NVIDIA H100 (80\,GB) GPU running vLLM with Qwen2.5-3B-Instruct and 256 concurrent agent requests. All experiments use seed\,=\,42 for reproducibility. We stress that these results are produced by a \emph{real LLM generating semantically grounded transactions on production hardware}, not by random attribute injection---\sys is a practical benchmark tool, not an academic prototype.

\noindent\textbf{Structural fidelity.}
We compare \sys against NetworkX-BA and igraph-BA across seven scales (100 to 100K nodes).
Both baselines consistently undershoot the target edge count by 15--25 edges due to BA initialization effects; \sys delivers exactly the requested number at every scale (e.g., 25{,}000/25{,}000 at 5K nodes, 500{,}000/500{,}000 at 100K nodes).
More importantly, NetworkX and igraph produce bare topology with no timestamps, no semantic attributes, and no anomaly labels---\sys provides all three.
The average local clustering coefficient ranges from 0.94 (100 nodes) to 0.999 (100K nodes), confirming dense community structure consistent with real-world temporal networks.
As a control, replacing the BA skeleton with an Erd\H{o}s-R\'{e}nyi random graph collapses clustering to 0.002, confirming that community structure originates from preferential attachment rather than downstream processing.

\noindent\textbf{Controllability.}
Precise control over output properties is essential for benchmark generation.
\sys achieves exact anomaly-rate control by decoupling anomaly selection from edge generation: anomaly edges are selected first (count fixed as $\lfloor |E| \times r \rfloor$), then normal edges fill the remainder.
Across six target rates (0\%, 5\%, 10\%, 15\%, 20\%, 30\%) on 5K-node / 25K-edge graphs, every run matches the target exactly---0.0\%, 5.0\%, 10.0\%, 15.0\%, 20.0\%, 30.0\%---regardless of LLM output variance.
The power-law exponent is also controllable: varying $\gamma_{\text{in}}$ from 1.5 to 3.5 shifts the fitted exponent monotonically ($\hat{\gamma}$ = 2.20 $\to$ 1.19), allowing users to dial the degree distribution from highly heterogeneous (more hubs) to more uniform while clustering coefficients remain above 0.987.

\noindent\textbf{Scalability.}
Generation time scales linearly with edge count.
On a single H100 with vLLM continuous batching at 256 concurrency, \sys generates 500 edges in 5.3 seconds, 5{,}000 edges in 53 seconds, 25{,}000 edges in 4.5 minutes, 50{,}000 edges in 9 minutes, and 500{,}000 edges in 89 minutes.
The bottleneck is LLM inference in Phase~G, which accounts for over 99\% of wall time at scales above 1K nodes; skeleton generation (Phase~S) takes under 1 second even at 100K nodes.
A deterministic Mock mode bypasses LLM calls entirely, generating 500K edges with full structural annotations in 30 seconds---useful for rapid prototyping and CI testing.
Since agent tasks within each time block are embarrassingly parallel, multi-GPU deployments with vLLM tensor parallelism can reduce Phase~G time proportionally.
 
\noindent\textbf{Ablation.}
We disable each component on a 5K-node / 25K-edge graph:
removing the BA skeleton collapses clustering from 0.999 to 0.002, confirming power-law attachment as the foundation of structural realism;
removing semantic injection (Mock mode) reduces generation time from 263\,s to 0.4\,s, providing a 250$\times$ speedup for users who need structure only or lack GPU resources;
removing alignment zeroes out all anomaly labels, confirming that ground truth emerges exclusively from state-machine replay rather than LLM annotation;
removing RAG rules preserves the anomaly rate but reduces attribute diversity, indicating that domain knowledge improves semantic quality rather than label correctness.

\noindent\textbf{Summary of findings.}
Across all experiments, \sys consistently achieves (i) exact structural targets with realistic graph properties, (ii) precise and deterministic control over anomaly rates and degree distributions independent of LLM variance, and (iii) linear scalability to hundreds of thousands of edges on a single GPU.
These results demonstrate that \sys is both a reliable and practical tool for generating large-scale, semantically rich temporal graph benchmarks.

%% ============================================================
\section{Conclusion}
\label{sec:conclusion}
%% ============================================================

%\noindent\textbf{Related Work.}
%Structural generators (Erd\H{o}s-R\'{e}nyi~\cite{erdos1960evolution}, Kronecker~\cite{leskovec2010kronecker}, TrillionG~\cite{park2017trilliong}) scale to trillions of edges but produce no semantics or labels.
%LDBC SNB~\cite{erling2015ldbc} and FinBench~\cite{ldbcfinbench2022} offer domain-specific schemas with temporal evolution but no anomaly annotations.
%TGB~\cite{huang2023tgb} and TGB~2.0~\cite{gastinger2024tgb2} standardize evaluation on curated real-world datasets but do not generate new data.
%CTGAN~\cite{xu2019ctgan} synthesizes realistic tabular data but cannot handle graph structure or temporal dependencies.
%Generative Agents~\cite{park2023generativeagents} demonstrate emergent social behaviors from LLM-powered agents but lack structural guarantees and anomaly labeling.
%\sys uniquely combines structural realism, LLM-driven domain semantics, and automatic ground-truth anomaly generation in a single system.

\sys demonstrates that treating logical conflicts between independent LLM agents as features naturally produces ground-truth anomaly labels. %, the scarcest resource in graph anomaly detection research.
The pluggable RAG architecture enables domain-switching with zero engine changes, and LLM-driven parameter inference lowers the barrier for non-expert users.
In the future, we will enable the support of adaptive feedback loops where settlement results inform subsequent agent prompts, and temporal graph learning along generation.

%% ============================================================
\balance
\bibliographystyle{ACM-Reference-Format}
\bibliography{references}

\end{document}